\documentclass[twocolumn,showpacs,preprintnumbers,amsmath,amssymb]{revtex4}

\usepackage{graphicx}
\usepackage{dcolumn}
\usepackage{bm}
\usepackage{amsmath,amsfonts,amscd,revsymb,latexsym, enumerate,multirow}
\newtheorem{theorem}{Theorem}[section]

\newtheorem{lemma}[theorem]{Lemma}

\newtheorem{proposition}[theorem]{Proposition}
\newtheorem{definition}[theorem]{Definition}

\newcommand{\qed}{{\hfill$\Box$}}
\newenvironment{proof}{\noindent \textbf{{Proof~} }}{\qed}

\def\bi{\begin{itemize}}
\def\ei{\end{itemize}}
\def\be{\begin{equation}}
\def\ee{\end{equation}}
\def\bea{\begin{eqnarray}}
\def\eea{\end{eqnarray}}
\def\ben{\begin{eqnarray*}}
\def\een{\end{eqnarray*}}

\def\>{\rangle}
\def\<{\langle}

\def\br{{\bf r}}

\newcommand{\bE}{{\mathbf E}}

\newcommand{\ba}{{\mathbf a}}
\newcommand{\bb}{{\mathbf b}}
\newcommand{\bc}{{\mathbf c}}
\newcommand{\bd}{{\mathbf d}}

\newcommand{\bee}{{\mathbf e}}

\def\bbZ{\mathbb{Z}}

\newcommand{\1} I 

\newcommand{\bra}[1]{\langle #1 |}
\newcommand{\ket}[1]{| #1 \rangle}

\newcommand{\proj}[1]{| #1 \>\!\< #1 |}

\def\*{\star}

\def\tilde{\widetilde}
\def\bar{\overline}

\def\cC{{\cal C}}
\def\cD{{\cal D}}        \def\cE{{\cal E}}

\def\cG{{\cal G}}        \def\cH{{\cal H}}

\def\cS{{\cal S}}        

\def\cV{{\cal V}}        

        \def\cZ{{\cal Z}}

\def\0{{\mathbf{0}}}
\def\1{{\mathbf{1}}}
\def\2{{\mathbf{2}}}
\def\3{{\mathbf{3}}}
\def\4{{\mathbf{4}}}
\def\5{{\mathbf{5}}}
\def\6{{\mathbf{6}}}
\def\7{{\mathbf{7}}}
\def\8{{\mathbf{8}}}
\def\9{{\mathbf{9}}}

\begin{document}

\title{General entanglement-assisted quantum error-correcting codes}

\author{Min-Hsiu Hsieh}
\email{minhsiuh@usc.edu}
\author{Igor Devetak}%
\email{devetak@usc.edu}
\author{Todd Brun}
\email{tbrun@usc.edu}
\affiliation{%
Ming Hsieh Department of Electrical Engineering, \\
University of Southern California, \\
Los Angeles, CA 90089 \\
}%

\date{\today}

\begin{abstract}
Entanglement-assisted quantum error-correcting codes (EAQECCs) make
use of pre-existing entanglement between the sender and receiver to
boost the rate of transmission.  It is possible to construct an
EAQECC from any classical linear code, unlike standard QECCs which
can only be constructed from dual-containing codes.  Operator
quantum error-correcting codes (OQECCs) allow certain errors to be
corrected (or prevented) {\it passively}, reducing the complexity of
the correction procedure.  We combine these two extensions of
standard quantum error correction into a unified
entanglement-assisted quantum error correction formalism. This new
scheme, which we call entanglement-assisted operator quantum error
correction (EAOQEC), is the most general and powerful quantum
error-correcting technique known, retaining the advantages of both
entanglement-assistance and passive correction. We present the
formalism, show the considerable freedom in constructing EAOQECCs
from classical codes, and demonstrate the construction with
examples.
\end{abstract}

\maketitle

\section{Introduction}

Conventional quantum error correcting
codes are simultaneous eigenspaces of a group of commuting
operators, the stabilizer group. A construction of Calderbank, Shor
and Steane \cite{CS96,Ste96} showed that it was possible to
construct quantum codes from classical binary codes--the CSS
codes--thereby drawing on the well-studied theory of classical error
correction. Later on, it was shown that \cite{CRSS98,DG97thesis} the
construction of quantum codes from classical codes can be put in a
more general framework, the stabilizer formalism.  This gave, among
other important benefits, a strong connection between quantum
error-correcting codes and classical symplectic codes, which are
closely related to linear quaternary codes (that is, linear codes
over $GF(4)$).

This connection between classical codes and quantum codes is
not universal, however. Rather, only classical codes that satisfy a
dual-containing constraint (i.e., that have self-orthogonal
parity-check matrices) can be used to construct standard quantum codes.
While this constraint is not too difficult to satisfy for
relatively small codes, it is a substantial barrier to the use of
highly efficient modern codes, such as Turbo codes and Low-Density Parity
Check (LDPC) codes, in quantum information theory.  These codes are
capable of achieving the classical capacity; but the difficulty of
constructing dual-containing versions of them has made progress
toward quantum versions very slow.

Recently, there have been two major breakthroughs in quantum error correction
theory. The first was the discovery of operator quantum
error-correcting codes (OQECCs)
\cite{AKS06a,Bacon05a,bacon-2006,KS06a,DRD05,Kribs06,NP05,DB05}.
These provide a general theory which combines passive error-avoiding
schemes, such as decoherence-free subspaces and noiseless subsystems,
with conventional (active) quantum error correction. In a certain sense, OQECC
does not lead to new codes, but instead provides a new kind of decoding
procedure: it is not necessary to actively correct all errors, but
rather only to perform correction modulo the subsystem structure.
One potential benefit of the new decoding procedure is to improve
the threshold of fault-tolerant quantum computation \cite{Bacon05a}.

The second breakthrough was the development of a theory of
entanglement-assisted quantum error correcting codes
\cite{Bow02,EAQECC2,BDH06}.  In this theory, it is assumed that
in addition to a quantum channel, the sender and receiver share
a certain amount of pre-existing entanglement.  The EAQECC formalism
can be applied to any classical quaternary code, not just
dual-containing ones, and the performance of the resulting quantum
code (that is, its minimum distance and net rate) is determined by
the performance of the classical code.
(OQECCs also allow quantum codes to be constructed from
classical codes which do not obey the dual-containing constraint,
but in this case the performance of the quantum codes cannot
be predicted from the performance of the classical codes).

Within the framework of EAQECCs, the existing theory of quantum error
becomes a special case in which the needed entanglement is zero.
Classical dual-containing codes give rise to standard quantum codes,
while all other classical codes give rise to EAQECCs.  In a similar way,
standard QECCs can also be thought of as a special of OQECCs,
where the protected subsystem is the entire system.  In this paper,
we move one step further, by incorporating both operator
quantum error correction and entanglement-assisted quantum
error correction into a single unified formalism. This unified scheme
is the most general theory of quantum error correction currently known.

We now briefly outline the structure of this paper.  In section II, we review
the construction of EAQECCs and OQECCs as extensions of the
usual stabilizer formalism.  In section III, we provide
the theoretical derivation of EAOQECCs, and briefly
discuss the relationship between conventional QECCs, OQECCs,
EAQECCs, and EAOQECCs. In section IV, we give some examples
of EAOQECCs, and show how one can make trade-offs between
entanglement-assistance and passive error correction.  Finally,
in section V we conclude.

\section{Review of EAQECCs and OQECCs}

First, let us recall the stabilizer formalism for conventional
quantum error-correcting codes. Let $\cG_n$ be the $n$-fold Pauli
Group \cite{NC00}.  Every operator in $\cG_n$ has either eigenvalues
$\pm1$ or $\pm i$. Let $\cS\subset\cG_n$ be an abelian subgroup
which does not contain $-I$.  Then this subgroup has a common
eigenspace $C(\cS)$ of $+1$ eigenvectors, which we call the {\it
code space} determined by the stabilizer $\cS$. Later on, we will
just use $C$ to denote the code space. Typically, the stabilizer is
represented by a minimal generating set $\{g_1,\ldots,g_m\}$, which
makes this a very compact way to specify a code (analogous to
specifying a classical linear code by its parity-check matrix).  We
write $\cS = \langle g_1,\ldots,g_m\rangle$ to denote that $\cS$ is
generated by $\{g_1,\ldots,g_m\}$.

Let $\bE\subset\cG_n$ be a set of possible errors. If a particular
error $E_1\in \bE$ anticommutes with any of the generators of $\cS$,
then the action of that error can be detected by measuring the
generators; if the measurement returns $-1$ instead of $1$, we know
an error has occurred.  On the other hand if the error is
actually {\it in} the stabilizer $\cS$, then it leaves all the
states in $C$ unchanged.  We can conclude that the code
$C$ can correct any error in $\bE$ if either $E_2^\dagger E_1 \notin
\cZ(\cS)$ or $E_2^\dagger E_1 \in \cS$ for all pairs of errors $E_1$ and $E_2$ in
$\bE$, where $\cZ(\cS)$ is the {\it centralizer} of $\cS$.

We can now generalize this description to the entanglement-assisted
case.  Given a \emph{nonabelian} subgroup ${\cal S}\subset{\cal
G}_n$ of size $2^m$, there  exists a set of generators
$\{\bar{Z}_1,\cdots,\bar{Z}_{s+c},\bar{X}_{s+1},\cdots,\bar{X}_{s+c}\}$
for $\cal S$ with the following commutation relations:
\begin{equation}\label{comm}
\begin{split}
[\bar{Z}_i,\bar{Z}_j]&=0 \ \ \ \ \forall i,j  \\
[\bar{X}_i,\bar{X}_j]&=0 \ \ \ \ \forall i,j  \\
[\bar{X}_i,\bar{Z}_j]&=0 \ \ \ \ \forall i\neq j \\
\{\bar{X}_i,\bar{Z}_i\}&= 0 \ \ \ \ \forall i.
\end{split}
\end{equation}
The parameters $s$ and $c$ satisfy $s+2c=m$. Let $\cS_I$ be the
\emph{isotropic} subgroup generated by
$\{\bar{Z}_{1},\cdots,\bar{Z}_{s}\}$ and $\cS_E$ be the
\emph{entanglement} subgroup generated by
$\{\bar{Z}_{s+1},\cdots,\bar{Z}_{s+c},\bar{X}_{s+1},\cdots,\bar{X}_{s+c}\}$.
The sizes of $\cS_I$ and $\cS_E$ describe the number of ancillas and
the number of ebits needed to construct EAQECCs, respectively.
(An ebit is one copy of a maximally entangled pair.)
The pair of subgroups $(\cS_I,\cS_E)$ defines an $[[n,k;c]]$ EAQECC
$C^{\text{ea}}$ that encodes $k=n-s-c$ logical qubits into $n$
physical qubits, with the help of $c$ ebits  shared between sender
and receiver and $s$ ancillas. These $n$ qubits are transmitted from
Alice (the sender) to Bob (the receiver), who measures them together
with his half of the $c$ ebits in order to correct any errors and
decode the $k$ logical qubits. We define $(k-c)/n$ as the net rate
of the code. This EAQECC $C^{\text{ea}}$ can correct an error set
$\bE$ if for all $E_1, E_2 \in \bE$, $E_2^\dagger E_1\in \cS_I
\bigcup (\cG_n-\cZ(\langle\cS_I,\cS_E\rangle))$.

The starting point for OQECCs is similar to that for EAQECCs.  Let
the nonabelian group $\cS\subset\cG_n$ of size $2^m$ be generated by
$\{\bar{Z}_1,\cdots,\bar{Z}_{s+r},\bar{X}_{s+1},\cdots,\bar{X}_{s+r}\}$,
where $\bar{Z}$'s and $\bar{X}$'s obey the same commutation
relations as in (\ref{comm}), and the parameters $s$ and $r$ satisfy
$s+2r=m$. Let $\cS_I=\langle \bar{Z}_{1},\cdots,\bar{Z}_{s} \rangle$
be the isotropic subgroup, and let $\cS_G=\langle
\bar{Z}_{s+1},\cdots,\bar{Z}_{s+r},\bar{X}_{s+1},\cdots,\bar{X}_{s+r}
\rangle$ be the \emph{gauge} subgroup. The size of $\cS_I$ and
$\cS_G$ describes the number of ancillas and the number of gauge
qubits (gauge qubits can be thought of as redundant logical qubits
to accommodate more errors) needed to construct OQECCs,
respectively. Then the pair of subgroups $(\cS_I,\cS_G)$ defines an $[[n,k;r]]$ OQECC
$C^{\text{op}}$ that fixes a $2^{r+k}$-dimensional code space, where
$s+k+r=n$. Furthermore, the gauge subgroup $\cS_G$ defines an
equivalence between pairs of states inside the code space: the two states
$\rho$ and $\rho'$ are considered to carry the same information
if they differ by the action of a quantum operation in the algebra generated
by $\cS_G$. These $r$ logical gauge qubits provide extra
power of passive error correction. This OQECC $C^{\text{op}}$ can
correct an error set $\bE$ if for all $E_1, E_2 \in \bE$,
$E_2^\dagger E_1\in \langle \cS_I,\cS_G \rangle \bigcup
(\cG_n-\cZ(\cS_I))$.

\section{Entanglement-assisted operator quantum error-correcting codes}
\subsection{The canonical code}
We illustrate the idea of EAOQECCs by the following canonical code.
Consider the trivial encoding operation $\cE_0$ defined by
\begin{equation}
\label{eaoqecc1} \cE_0:\proj{\psi} \to \proj{\0} \otimes \proj{\Phi}
\otimes \sigma \otimes \proj{\psi} .
\end{equation}
The operation simply appends $s$ ancilla qubits in the state
$\ket{\0}$, $c$ copies of $\ket{\Phi}$ (a maximally entangled
state shared between sender Alice and receiver Bob), and an
arbitrary state $\sigma$ of size $r$ qubits, to the initial register
containing the state $\ket{\psi}$ of size $k$ qubits, where $s+k+r+c=n$. These
$r$ extra qubits are the gauge qubits.  Two states of this form which differ
only in $\sigma$ are considered to encode the same quantum information.

\begin{proposition}
\label{code1} The encoding given by $\cE_0$ and a suitably-defined
decoding map $\cD_0$ can correct the error set
\be
\begin{split}
\bE_0 =& \{X^\ba Z^\bb  \otimes Z^{\ba_1}X^{\ba_2} \otimes X^\bc
Z^\bd \otimes X^{\alpha(\ba,\ba_1,\ba_2)} Z^{\beta(\ba,\ba_1,\ba_2)}
:
\\ & \ba,\bb \in (\bbZ_2)^s, \ba_1,\ba_2 \in (\bbZ_2)^c, \bc,\bd \in (\bbZ_2)^r \} ,
\end{split}
\ee
for any fixed functions $\alpha,\beta:(\bbZ_2)^s \times
(\bbZ_2)^c \times (\bbZ_2)^c \to (\bbZ_2)^k$.
\end{proposition}

\begin{proof}
After applying an error $E \in \bE_0$, the channel output becomes
(up to a phase factor): \begin{equation}
\begin{split}
(X^\ba Z^\bb)\proj{\0} &(X^\ba Z^\bb)^\dagger \otimes \\
(Z^{\ba_1}X^{\ba_2}\otimes I^B) \proj{\Phi}&
(Z^{\ba_1}X^{\ba_2}\otimes I^B)^\dagger  \otimes \\
(X^\bc Z^\bd) \sigma &(X^\bc Z^\bd)^\dagger \otimes \\
(X^{\alpha(\ba,\ba_1,\ba_2)} Z^{\beta(\ba,\ba_1,\ba_2)})
\proj{\psi} &(X^{\alpha(\ba,\ba_1,\ba_2)} Z^{\beta(\ba,\ba_1,\ba_2)})^\dagger \\
= \proj{\ba} \otimes \proj{\ba_1,\ba_2} &\otimes \sigma' \otimes \proj{\psi'}
\end{split}
\end{equation} where $\ket{\ba}=X^{\ba}\ket{\0}$, $\ket{\ba_1,\ba_2} =
(Z^{\ba_1} X^{\ba_2}\otimes I^B) \ket{\Phi}^{\otimes c}$,
$\sigma'=(X^\bc Z^\bd) \sigma (X^\bc Z^\bd)^\dagger$, and
$\ket{\psi'}=(X^{\alpha(\ba,\ba_1,\ba_2)}
Z^{\beta(\ba,\ba_1,\ba_2)}) \ket{\psi}$. Here we write, e.g.,
\[
X^{\ba} \equiv X^{a_1}\otimes X^{a_2}\otimes\cdots X^{a_s},
\]
where $\ba=(a_1,\ldots,a_s)\in(\bbZ_2)^s$, $X^0 = I$, and $X^1 = X$.
As the vector $(\ba, \ba_1, \ba_2, \bb, \bc, \bd)$ completely
specifies the error operator $E$, it is called the \emph{error
syndrome}. However, in order to correct this error, only the
\emph{reduced syndrome} $(\ba,\ba_1,\ba_2)$ matters. Here two kinds
of passive error correction are involved. The errors that come from
vector $\bb$ are passively corrected because they do not affect the
encoded state given in Eq.~(\ref{eaoqecc1}). The errors that come
from vector $(\bc,\bd)$ are passively corrected because of the
subsystem structure inside the code space: $\rho\otimes\sigma$ and
$\rho\otimes\sigma'$ represent the same information, differing only
by a gauge operation.

The decoding operation $\cD_0$ is constructed based on the reduced
syndrome, and is also known as \emph{collective measurement}. Bob
can recover the state $\ket{\psi}$ by performing the decoding
$\cD_0$: \be\label{decode}
\begin{split}
\cD_0 = &\sum_{\ba,\ba_1,\ba_2} \ket{\ba}\bra{\ba}\otimes
\ket{\ba_1,\ba_2}\bra{\ba_1,\ba_2} \otimes I \\ &\otimes
X^{-\alpha(\ba,\ba_1,\ba_2)}Z^{-\beta(\ba,\ba_1,\ba_2)},
\end{split}
\ee followed by discarding the unwanted systems.
\end{proof}

We can rephrase the above error-correcting procedure in terms of the
stabilizer formalism. Let $\cS_0=\langle
\cS_{0,I},\cS_{0,S}\rangle$, where $\cS_{0,I}=\langle Z_1 , \cdots ,
Z_s \rangle$ is the isotropic subgroup of size $2^s$ and
$\cS_{0,S}=\langle
Z_{s+1},\cdots,Z_{s+c+r},X_{s+1},\cdots,X_{s+c+r}\rangle$ is the
\emph{symplectic} subgroup of size $2^{2(c+r)}$. We can further
divide the symplectic subgroup $\cS_{0,S}$ into an entanglement
subgroup
\[
\cS_{0,E}=\langle Z_{s+1},\cdots,Z_{s+c},X_{s+1},\cdots,X_{s+c} \rangle
\]
of size $2^{2c}$ and a gauge subgroup
\[
\cS_{0,G}=\langle Z_{s+c+1},\cdots,Z_{s+c+r},X_{s+c+1},\cdots,X_{s+c+r} \rangle
\]
of size $2^{2r}$, respectively. The generators of $(\cS_{0,I},\cS_{0,E},\cS_{0,G})$
are arranged in the following form:
\begin{equation}
\begin{array}{cccc}
Z^{\bee_i} & I & I & I \\
I & Z^{\bee_j} & I & I \\
I & X^{\bee_j} & I & I \\
I & I & Z^{\bee_l} & I \\
I & I & X^{\bee_l} & I \\
\overleftrightarrow{s} & \overleftrightarrow{c} &
\overleftrightarrow{r} & \overleftrightarrow{k}
\end{array}
\end{equation}
where $\{\bee_i\}_{i\in[s]}$, $\{\bee_j\}_{j\in[c]}$, and
$\{\bee_l\}_{l\in[r]}$ are the set of standard bases in
$(\bbZ_2)^s$, $(\bbZ_2)^c$, and $(\bbZ_2)^r$, respectively, and
$[k]\equiv\{1,\cdots,k\}$.

It follows that the three subgroups $(\cS_{0,I},\cS_{0,E},\cS_{0,G})$ define the
canonical EAOQECC given in (\ref{eaoqecc1}). The
subgroups $\cS_{0,I}$ and $\cS_{0,E}$ define a $2^{k+r}$-dimensional
code space $C_0^{\text{eao}}\subset \cH^{\otimes (n+c)}$, and the
gauge subgroup $\cS_{0,G}$ specifies all possible operations that
can happen on the gauge qubits. Thus we can use $\cS_{0,G}$ to
define an equivalence class between two states in the code space of
the form: $\rho\otimes \sigma$ and $\rho\otimes\sigma'$, where
$\rho$ is a state on $\cH^{\otimes k}$, and $\sigma,\sigma'$ are states on $\cH^{\otimes r}$.
Consider the parameters of the canonical code. The number of
ancillas $s$ is equal to the number of generators for the isotropic
subgroup $\cS_{0,I}$. The number of ebits $c$ is equal to the number
of symplectic pairs that generate the entanglement subgroup $\cS_{0,E}$. The
number of gauge qubits $r$ is equal to the number of symplectic
pairs for the gauge subgroup $\cS_{0,G}$. Finally, the number of
logical qubits $k$ that can be encoded in $C_0^{\text{eao}}$ is
equal to $n-s-c-r$. To sum up, $C_0^{\text{eao}}$ defined by
$(\cS_{0,I},\cS_{0,E},\cS_{0,G})$ is an $[[n,k;r,c]]$ EAOQECC that
fixes a $2^{k+r}$-dimensional code space, within which $\rho\otimes
\sigma$ and $\rho\otimes \sigma'$ are considered to carry the same
information. Notice that there is a tradeoff between the number of
encoded bits and gauge bits, in that we can reduce the rate by improving
the error-avoiding ability or vice versa.

\begin{proposition}
\label{dumb} The EAOQECC $C_0^{\text{eao}}$ defined by
$(\cS_{0,I},\cS_{0,E},\cS_{0,G})$ can correct an error set $\bE_0$
if for all $E_1,E_2\in \bE_0$, $E_2^\dagger E_1 \in
\langle\cS_{0,I},\cS_{0,G}\rangle\bigcup (\cG_n-\cZ(\langle
\cS_{0,I},\cS_{0,E} \rangle))$.
\end{proposition}
\begin{proof}
Since the vector $(\ba,\ba_1,\ba_2,\bb,\bc,\bd)$ completely
specifies the error operator $E$, we consider the following two
different cases:
\begin{itemize}
\item If two error operators $E_1$ and $E_2$ have the same reduced
syndrome $(\ba,\ba_1,\ba_2)$, then the error operator $E_2^\dagger
E_1$ gives us all-zero reduced syndrome with some vector
$(\bb,\bc,\bd)$. Therefore, $E_2^\dagger
E_1\in\langle\cS_{0,I},\cS_{0,G}\rangle$. This error $E_2^\dagger
E_1$ has no effect on the logical state $\proj{\psi}$.
\item If two error operators $E_1$ and $E_2$ have different reduced
syndromes, and let $(\ba,\ba_1,\ba_2)$ be the reduced syndrome of
$E_2^\dagger E_1$, then $E_2^\dagger E_1 \not\in Z(\langle
\cS_{0,I},\cS_{0,E}\rangle)$. This error $E_2^\dagger E_1$ can be
corrected by the decoding operation given in (\ref{decode}).
\end{itemize}
\end{proof}

\subsection{The general case}
Before giving the theorem, we first state two lemmas that lead
directly to the result.
\begin{lemma}
\label{basis} Let $\cV$ be an arbitrary subgroup of $\cG_n$ with
size $2^m$. Then there exists a set of generators
$\{\bar{Z}_1,\cdots,\bar{Z}_{p+q},\bar{X}_{p+1},\cdots,\bar{X}_{p+q}\}$
that generates $\cV$ such that $\bar{Z}$'s and $\bar{X}$'s obey the
same commutation relations as in (\ref{comm}), for some $p,q\geq 0$
and $p+2q=m$.
\end{lemma}
\begin{proof}
See \cite{EAQECC2}.
\end{proof}

Consider an arbitrary nonabelian group $\cS$ of size
$2^{s+2(c+r)}$, for some $s,c,r\geq 0$, lemma \ref{basis} says that
there exists a set of generators
$\{\bar{Z}_{1},\cdots,\bar{Z}_{s+c+r},\bar{X}_{s+1},\cdots,\bar{X}_{s+c+r}\}$
such that $\cS=\langle\cS_I,\cS_S\rangle$, where
$\cS_I=\langle\bar{Z}_{1},\cdots,\bar{Z}_{s}\rangle$ is the
isotropic subgroup, and
$\cS_S=\langle\bar{Z}_{s+1},\cdots,\bar{Z}_{s+c+r},\bar{X}_{s+1},\cdots,\bar{X}_{s+c+r}\rangle$
is the symplectic subgroup. Furthermore, the symplectic subgroup
$\cS_S$ can be divided into the entanglement subgroup $\cS_E$ of
size $2^{2c}$ and the gauge subgroup $\cS_G$ of size $2^{2r}$.

\begin{lemma}
\label{sim} If there is a one-to-one map between $\cV$ and $\cS$
which preserves their commutation relations, which we denote
$\cV\sim\cS$, then there exists a unitary $U$ such that for each
$V_i\in\cV$, there is a corresponding $S_i\in\cS$ such that $V_i=U
S_i U^{-1}$, up to a phase which can differ for each generator.
\end{lemma}
\begin{proof}
See \cite{EAQECC2}.
\end{proof}

This lemma enables us to link the group $\cS$ to $\cS_0$ (in other
words, map $(\cS_I,\cS_E,\cS_G)$ to
$(\cS_{0,I},\cS_{0,E},\cS_{0,G})$) by some unitary $U$ such that \be
\begin{split}
Z_i &= U \bar{Z}_i U^{-1}, \forall i\in\{1,2,\cdots,s+c+r\}   \\
X_j &= U \bar{X}_j U^{-1}, \forall j\in\{s+1,\cdots,s+c+r\}.
\end{split}
\ee Let $U$ also denote the trivial extension of $U$ that acts as the identity
on the qubits on Bob's side. We can now define an $[[n,k;r,c]]$ EAOQECC
$C^{\text{eao}}$ by $(\cS_I,\cS_S,\cS_G)$, that incorporates both
entanglement-assistance and passive error avoiding ability.

We now reach our main theorem in this paper:
\begin{theorem}
\label{general} Given the subgroups $(\cS_I,\cS_E,\cS_G)$, there exists an
$[[n,k;r,c]]$ entanglement-assisted operator quantum
error-correcting code $C^{\text{eao}}$ defined by the encoding and
decoding pair: $(\cE,\cD)$. The code $C^{\text{eao}}$ can correct
the error set $\bE$ if for all $E_1, E_2 \in \bE$, $E_2^\dagger E_1
\in \langle\cS_I,\cS_G\rangle \bigcup
(\cG_n-\cZ(\langle\cS_I,\cS_E\rangle))$.
\end{theorem}

\begin{proof}
Since $\cS\sim\cS_0$, there exists an unitary matrix $U$ that
preserves the commutation relations. Define $\cE=U^{-1}\circ\cE_0$ and
$\cD=\cD_0\circ U$, where $\cE_0$ and $\cD_0$ are given in
(\ref{eaoqecc1}) and (\ref{decode}), respectivley. Since
\[
\cD_0\circ E_0\circ \cE_{0}=\text{id}^{\otimes k}
\]
for any $E_0\in\bE_0$, then
\[
\cD\circ E\circ\cE=\text{id}^{\otimes k}
\]
follows for any $E\in\bE$. Thus, the encoding and decoding pair $(\cE,\cD)$ corrects
$\bE$.
\end{proof}

\subsection{Properties of EAOQECCs}
Conventionally, the performance of a code is characterized by its
distance $d$. Define the \emph{weight} of a Pauli operator to be the
number of single qubit operators that are not the identity. We say
that the $[[n,k,d;r,c]]$ EAOQECC $C^{\text{eao}}$ has distance $d$
if it can correct any error set $\bE$ such that for each operator
$E\in \bE$, the weight $t$ of $E$ satisfies $2t+1\leq d$.

In the description earlier in this section, we assumed that the
gauge subgroup was generated by a set of symplectic pairs of
generators.  In some cases, it may make sense to start with a
gauge subgroup which itself has both an isotropic (i.e., commuting)
and a symplectic subgroup.  In this case, we can arbitrarily add
a symplectic partner for each generator in the isotropic subgroup
of the gauge group.  This can be useful in constructing EAOQECCs
from EAQECCs, in a way analogous to how OQECCs can
be constructed by starting from standard QECCs.  Poulin shows in \cite{DB05} that it is
possible to move generators from the stabilizer group into the gauge subgroup,
together with their symplectic partners, without changing the
essential features of the original code. We provide an example of such
a construction in section \ref{DBexample}.

There is further flexibility in trading between active
error correction ability and passive noise avoiding ability
\cite{AKS06a}. This is captured by the following theorem:

\begin{theorem}
\label{trans} We can transform any $[[n,k+r,d_1;0,c]]$ code $C_1$
into an $[[n,k,d_2;r,c]]$ code $C_2$, and transform the
$[[n,k,d_2;r,c]]$ code $C_2$ into an $[[n,k,d_3;0,c]]$ code $C_3$,
where $d_1\leq d_2 \leq d_3$.
\end{theorem}
\begin{proof}
There exists an isotropic subgroup $\cS_I$ and an entanglement subgroup
$\cS_E$ associated with $C_1$ of size $2^{s}$ and $2^{2c}$,
respectively. These parameters satisfy $s+c+k+r=n$. This code $C_1$
corresponds to an $[[n,k+r,d_1;0,c]]$ EAQECC for some $d_1$. If we
add the gauge subgroup $\cS_G$ of size $2^{2r}$, then
$(\cS_I,\cS_E,\cS_G)$ defines an $[[n,k,d_2;r,c]]$ EAOQECC $C_2$ for
some $d_2$, which follows from theorem \ref{general}. Let $\bE_1$ be the
error set that can be corrected by $\cC_1$, and $\bE_2$ be the error
set that can be corrected by $\cC_2$. Clearly, $\bE_1\subset \bE_2$
(see the following table), so $\cC_2$ can correct more errors than
$\cC_1$. By sacrificing part of the transmission rate, we have gained
additional passive correction, and $d_2\geq d_1$.

If we now throw away half of each symplectic pair in $\cS_G$ and include
the remaining generators in $\cS_I$, which becomes $\cS_I'$,
the size of the isotropic subgroup increases by a factor of $2^r$. Then
$(\cS_I',\cS_E)$ defines an $[[n,k,d_3;0,c]]$ EAQECC $C_3$. Let
$\bE_3$ be the error set that can be corrected by $C_3$. Let
$E\in\bE_2$, then either $E\in\langle \cS_I,\cS_G\rangle$ or
$E\not\in\cZ(\langle \cS_I,\cS_E\rangle)$.
\begin{itemize}
\item If $E\in\langle \cS_I,\cS_G\rangle$, then either $E\in \cS_I'$
or $E\in\langle \cS_I,\cS_G\rangle/\cS_I'$. If $E\in\langle
\cS_I,\cS_G\rangle/\cS_I'$, this implies $E\not\in\cZ(\cS_I')$.
Thus, $E\in\bE_3$.
\item Since $\langle \cS_I,\cS_E\rangle\subset \langle
\cS_I',\cS_E\rangle$, we have $\cZ(\langle
\cS_I',\cS_E\rangle)\subset\cZ(\langle \cS_I,\cS_E\rangle)$. If
$E\not\in\cZ(\langle \cS_I,\cS_E\rangle)$, then
$E\not\in\cZ(\langle\cS_I',\cS_E \rangle)$. Thus, $E\in\bE_3$.
\end{itemize}
Putting these together we get $ \bE_2\subset \bE_3$. Therefore $d_3\geq d_2$.
\end{proof}

To conclude this section, we list the different error-correcting
criteria of a conventional stabilizer code (QECC), an EAQECC,
an OQECC, and an EAOQECC:
\begin{equation}
\begin{tabular}{|c|c|}
\hline QECC & EAQECC \\ \hline
$E_2^\dagger E_1\not\in \cZ(\cS_I)$ & $E_2^\dagger E_1\not\in \cZ(\langle\cS_I,\cS_E\rangle)$ \\
$E_2^\dagger E_1\in\cS_I$ & $E_2^\dagger E_1\in\cS_I$ \\ \hline
OQECC & EAOQECC \\ \hline
$E_2^\dagger E_1\not\in \cZ(\cS_I)$ & $E_2^\dagger E_1\not\in \cZ(\langle\cS_I,\cS_E\rangle)$ \\
$E_2^\dagger E_1\in \langle\cS_I,\cS_G\rangle$ & $E_2^\dagger E_1\in\langle\cS_I, \cS_G\rangle$ \\
\hline
\end{tabular} \nonumber
\end{equation}

\section{Examples}

\subsection{EAOQECC from EAQECC}
\label{DBexample}

Our first example constructs an $[[8,1,3;r=2,c=1]]$ EAOQECC from an
[[8,1,3;1]] EAQECC.  Consider the EAQECC code defined by the group $\cS$
generated by the operators in Table~\ref{EAQECC1}.
Here $\bar{Z}$ and $\bar{X}$ refer to the logical $Z$ and $X$
operation on the codeword, respectively. The isotropic subgroup is $\cS_I=\langle
S_1,S_2,S_3,S_4,S_5,S_8\rangle$, the entanglement subgroup is
$\cS_E=\langle S_6,S_7 \rangle$, and together they generate the full group
$\cS=\langle \cS_I,\cS_E\rangle$. This code $C(\cS_I,\cS_E)$ encodes one
qubit into eight physical qubits with the help of one ebit,
and therefore is an $[[8,1;1]]$ code. It can be easily checked that this
code can correct an arbitrary single-qubit error, and it is
degenerate.

\begin{table}[htdp]
\begin{center}
\begin{tabular}{c|cccccccc|c}
  & \multicolumn{8}{c} {Alice} & Bob        \\ \hline\hline
 $S_1$ & Z & Z & I & I & I & I & I & I & I\\
 $S_2$ & Z & I & Z & I & I & I & I & I & I\\
 $S_3$ & I & I & I & Z & Z & I & I & I & I\\
 $S_4$ & I & I & I & Z & I & Z & I & I & I\\
 $S_5$ & I & I & I & I & I & I & Z & Z & I\\
 $S_6$ & I & I & I & I & I & I & I & Z & Z\\
 $S_7$ & X & X & X & I & I & I & X & X & X\\
 $S_8$ & X & X & X & X & X & X & I & I & I\\ \hline
 $\bar{Z}$ & Z & I & I & Z & I & I & I & Z & I\\
 $\bar{X}$ & I & I & I & X & X & X & I & I & I\\
 \hline\hline
\end{tabular}
\end{center}
\caption{The original [[8,1,3;$c=1$]] EAQECC.}
\label{EAQECC1}
\end{table}%

By inspecting the group structure of $\cS$, we can recombine the first
four stabilizers of the code to give two isotropic generators (which we
retain in $\cS_I$), and two generators which we include, together with
their symplectic partners, in the subgroup $\cS_G$, for two
qubits of gauge symmetry. This yields an $[[8,1,3;2,1]]$ EAOQECC
whose generators are given in Table~\ref{EAOQECC1}.
where $\cS_I=\langle S_1',S_2',S_3',S_6'\rangle$, $\cS_E=\langle
S_4',S_5'\rangle$, and $\cS_G=\langle g_1^z,g_1^x,g_2^z,g_2^x\rangle$.

\begin{table}[htdp]
\begin{center}
\begin{tabular}{c|cccccccc|c}
  & \multicolumn{8}{c} {Alice} & Bob        \\ \hline\hline
 $S_1'$ & Z & Z & I & Z & Z & I & I & I & I\\
 $S_2'$ & Z & I & Z & Z & I & Z & I & I & I\\
 $S_3'$ & I & I & I & I & I & I & Z & Z & I\\
 $S_4'$ & I & I & I & I & I & I & I & Z & Z\\
 $S_5'$ & X & X & X & I & I & I & X & X & X\\
 $S_6'$ & X & X & X & X & X & X & I & I & I\\ \hline
 $\bar{Z}$ & Z & I & I & Z & I & I & I & Z & I\\
 $\bar{X}$ & I & I & I & X & X & X & I & I & I\\ \hline
 $g_1^z$ & Z & Z & I & I & I & I & I & I & I\\
 $g_1^x$ & I & X & I & I & X & I & I & I & I\\
 $g_2^z$ & I & I & I & Z & I & Z & I & I & I\\
 $g_2^x$ & I & I & X & I & I & X & I & I & I\\
 \hline\hline
\end{tabular}
\end{center}
\caption{The resulting [[8,1,3;$c=2$,$r=1$]] EAOQECC.}
\label{EAOQECC1}
\end{table}%

\subsection{EAOQECCs from classical BCH codes}

EAOQECCs can also be constructed directly from classical binary codes.
Before we give examples, however, we need one more theorem:

\begin{theorem}
\label{ebit} Let $H$ be any binary parity check matrix with
dimension $(n-k)\times n$. We can obtain the corresponding
$[[n,2k-n+c;c]]$ EAQECC, where $c = {\rm rank}(H H^T)$ is the number
of ebits needed.
\end{theorem}
\begin{proof}By the CSS construction,
let $\tilde{H}$ be \be \label{h4} \tilde{H}=\left(\begin{array}{c|c}
 H & \mathbf{0} \\ \mathbf{0} & H
\end{array}\right).
\ee Let $\cS$ be the group generated by $\tilde{H}$, then
$\cS=\langle Z^{\br_1},\cdots, Z^{\br_{n-k}},X^{\br_{1}},\cdots,X^{\br_{n-k}} \rangle$,
where $\br_i$ is the $i$-th row vector of $H$. Now we need to determine how many
symplectic pairs are in group $\cS$. Since rank$(H H^T)=c$, there
exists a matrix $P$ such that
\[
P H H^T P^T=\left(\begin{array}{cccc} I_{p\times p} & \0  & \0 & \0 \\ \0  &
  \mathbf{0} & I_{q\times q} & \0 \\ \0 &    I_{q\times q} & \mathbf{0}
  & \0
  \\ \0 & \0 & \0 & \mathbf{0}
\end{array}\right)_{(n-k)\times(n-k)}
\]
where $p+2q=c$. Let $\br_i'$ be the $i$-th row vector of the new
matrix $P H$, then $\cS=\langle Z^{\br_1'},\cdots, Z^{\br_{n-k}'}
,X^{\br_{1}'},\cdots,X^{\br_{n-k}'}\rangle $.

Using the fact that $\{Z^{\ba},X^{\bb}\}=0$ if and only if $\ba\cdot
\bb=1$, we know that the operators $Z^{\br_i'},X^{\br_i'}$ for $1\le i\le p$,
and the operators $Z^{\br_{p+j}'},X^{\br_{p+q+j}'}$ for $1 \le j\le q$,
generate a symplectic subgroup in $\cS$ of size $2^{2c}$.
\end{proof}

\label{BCHexample}
\begin{definition}\cite{FJM77}
A cyclic code of length $n$ over \text{GF}($p^m$) is a \emph{BCH
code of designed distance $d$} if, for some number $b\geq 0,$ the
generator polynomial $g(x)$ is
\[
g(x)=\text{lcm}\{M^{b}(x),M^{b+1}(x),\cdots,M^{b+d-2}(x)\},
\]
where $M^{k}(x)$ is the minimal polynomial of $\alpha^k$ over
GF($p^m$). I.e. $g(x)$ is the lowest degree monic polynomial over
GF($p^m$) having $\alpha^b,\alpha^{b+1},\cdots,\alpha^{b+d-2}$ as
zeros. When $b=1$, we call such BCH codes \emph{narrow-sense} BCH
codes. When $n=p^m-1$, we call such BCH codes \emph{primitive}.
\end{definition}

Consider the primitive narrow-sense BCH code over GF($2^6$). This
code has the following parity check matrix \be \label{BCH}
H_q=\left(\begin{array}{ccccc}
1 & \alpha & \alpha^2 & \cdots & \alpha^{n-1} \\
1 & \alpha^3 & \alpha^6 & \cdots & \alpha^{3(n-1)} \\
1 & \alpha^5 & \alpha^{10} & \cdots & \alpha^{5(n-1)} \\
1 & \alpha^7 & \alpha^{14} & \cdots & \alpha^{7(n-1)}
\end{array}\right),
\ee where $\alpha\in \text{GF}(2^6)$ satisfies $\alpha^6+\alpha+1=0$
and $n=63$. Since all finite fields of order $p^m$ are
\emph{isomorphic}, there exists a one-to-one correspondence between
elements in $\{\alpha^j:j=0,1,\cdots,p^m-2,\infty\}$ and elements in
$\{a_0a_1\cdots,a_m: a_i\in\text{GF}(p)\}$. If we replace
$\alpha^{j}\in\text{GF}(2^6)$ in (\ref{BCH}) with its binary
representation, this gives us a binary $[63,39,9]$ BCH code whose
parity check matrix $H_2$ is of size $24 \times 63$. If we carefully
inspect the binary parity check matrix $H_2$, we will find that the
first 18 rows of $H_2$ give a $[63,45,7]$ dual-containing BCH code.

From Theorem \ref{ebit}, it is easy to check that $c=\text{rank}(H_2
H_2^T)=6$. Thus by the CSS construction \cite{BDH06}, this binary
$[63,39,9]$ BCH code will give us a corresponding $[[63,21,9;6]]$
EAQECC.

If we further explore the group structure of this EAQECC, we will
find that the 6 symplectic pairs that generate the entanglement
subgroup $\cS_E$ come from the last 6 rows of $H_2$.  (Remember that
we are using the CSS construction.) If we remove one symplectic pair
at a time from $\cS_E$ and adding it to the gauge subgroup $\cS_G$,
we get EAOQECCs with parameters given in Table~\ref{BCHtable}.

\begin{table}[htdp]
\begin{center}
\begin{tabular}{ccccc}
 n & k & d & r & c \\ \hline
63 & 21 & 9 & 0 & 6 \\
63 & 21 & 7 & 1 & 5 \\
63 & 21 & 7 & 2 & 4 \\
63 & 21 & 7 & 3 & 3 \\
63 & 21 & 7 & 4 & 2 \\
63 & 21 & 7 & 5 & 1 \\
63 & 21 & 7 & 6 & 0
\end{tabular}
\end{center}
\caption{Parameters of the EAOQECCs constructed from a classical [63,39,9] BCH code.}
\label{BCHtable}
\end{table}%

In general, there could be considerable freedom in which of
the symplectic pairs is to be removed.
There are plenty of choices in the generators of $\cS_E$. In
fact, it does not matter which symplectic pair we remove first in
this example, due to the algebraic structure of this BCH code. The
distance is always lower bounded by $7$.

One final remark: this example gives EAOQECCs with positive net
rate, so they could be used as catalytic codes.

\subsection{EAOQECCs from classical quaternary codes}
\label{qexample}

In the following, we will show how to use MAGMA
\cite{MAGMA} to construct EAOQECCs from classical quaternary codes
with positive net yield and without too much distance degradation.
Consider the following parity check matrix $H_4$ of a $[15,10,4]$
quaternary code:
\begin{widetext}
\begin{equation} \label{QC}
H_4=\left(\begin{array}{ccccccccccccccc}
1&0&0&0&1&1&\omega^2&0&1&\omega^2&0&\omega&\omega^2&1&0 \\
 0&1&0&0&1&0&\omega&\omega^2&1&\omega&0&0&1&\omega&1 \\
 0&0&1&0&\omega&\omega^2&1&\omega&1&0&0&\omega&1&\omega^2&\omega \\
 0&0&0&1&1&\omega^2&0&1&\omega^2&\omega&0&\omega^2&1&0 &\omega^2 \\
 0&0&0&0&0&0&0&0&0&0&1&0&0&0&0 \\
\end{array}\right),
\end{equation}
\end{widetext}
where $\{0,1,\omega,\omega^2\}$ are elements of GF(4) that satisfy:
$1+\omega+\omega^2=0$ and $\omega^3=1$.  This quaternary code has
the largest minimum weight among all known $[n=15,k=10]$ linear
quaternary codes. By the construction given in \cite{BDH06}, this
code gives a corresponding $[[15,9,4;c=4]]$ EAQECC with the
stabilizers given in Table~\ref{EAQECC2}.

\begin{table}[htdp]
\begin{center}
\begin{tabular}{|c|ccccccccccccccc|}
\hline\hline \multirow{8}{*}{${\cal{S}}_E$} &I&I&Y&I&Z&X&Y&Z&Y&I&I&Z&Y&X&Z \\
&I&Y&I&I&Y&I&Z&X&Y&Z&I&I&Y&Z&Y \\
&I&Z&Y&I&I&X&Z&X&X&X&I&Z&X&I&I \\
&I&I&X&I&Y&Z&X&Y&X&I&I&Y&X&Z&Y \\
&I&I&I&I&I&I&I&I&I&I&Z&I&I&I&I \\
&I&I&I&I&I&I&I&I&I&I&Y&I&I&I&I \\
&I&Z&Z&Z&X&I&Y&I&Y&I&I&Z&Z&Z&I \\
&I&Y&Y&Y&Z&I&X&I&X&I&I&Y&Y&Y&I \\ \hline
\multirow{2}{*}{${\cal{S}}_I$}&Z&Z&Y&I&Z&Y&X&X&Y&Z&I&Y&Z&Z&I \\
&Y&Y&X&I&Y&X&Z&Z&X&Y&I&X&Y&Y&I\\
 \hline\hline
\end{tabular}
\end{center}
\caption{Stabilizer generators of the [[15,9,4;$c=4$]] EAQECC derived from the classical code given by Eq.~(\ref{QC}).}
\label{EAQECC2}
\end{table}

The entanglement subgroup $\cS_E$ of this EAQECC has $c=4$ symplectic
pairs. Our goal is to construct an EAOQECC from this EAQECC such
that the power of error correction is largely retained, but the amount
of entanglement needed is reduced. In this example, the choice of
which symplectic pair is removed strongly affects the distance $d$
of the resulting EAOQECC. By using MAGMA to perform a random search
of all the possible sympletic pairs in $\cS_E$, and then putting them into the
gauge subgroup $\cS_G$, we can obtain a $[[15,9,3;c=3,r=1]]$ EAOQECC
with stabilizers given in Table~\ref{EAOQECC2}.
The distance is reduced by one, which still retains the ability to correct all
one-qubit errors; the amount of entanglement needed is reduced by one ebit;
and we gain some extra power of passive error correction, due to the subsystem
structure inside the code space, given by the gauge subgroup $\cS_G$.

\section{Conclusion}

We have shown a very general quantum error correction scheme that
combines two extensions of standard stabilizer codes. This
scheme includes the advantages of both entanglement-assisted and
operator quantum error correction.

In addition to presenting the formal theory of EAOQECCs,
we have given several examples of code construction. The
methods of constructing OQECCs from standard QECCs can be applied
directly to the construction of EAOQECCs from EAQECCs.  We can
also construct EAOQECCs directly from classical linear codes.

We also show that, by exploring the structure of the symplectic
subgroup, we can construct versatile classes EAOQECCs with varying
powers of passive versus active error correction.  Starting with good
classical codes, this entanglement-assisted operator formalism can be
used to construct quantum codes tailored to the needs of particular
applications.  The study of such classes of good quantum codes is the
subject of ongoing research.

\begin{table}[htdp]
\begin{center}
\begin{tabular}{|c|ccccccccccccccc|}
\hline\hline \multirow{6}{*}{${\cal{S}}_E$} &I&I&Y&I&Z&X&Y&Z&Y&I&I&Z&Y&X&Z \\
&I&Y&I&I&Y&I&Z&X&Y&Z&I&I&Y&Z&Y \\
&I&Z&Y&I&I&X&Z&X&X&X&I&Z&X&I&I \\
&I&I&X&I&Y&Z&X&Y&X&I&I&Y&X&Z&Y \\
&I&I&I&I&I&I&I&I&I&I&Z&I&I&I&I \\
&I&I&I&I&I&I&I&I&I&I&Y&I&I&I&I \\
\hline \multirow{2}{*}{${\cal{S}}_G$}&I&Z&Z&Z&X&I&Y&I&Y&I&I&Z&Z&Z&I \\
&I&Y&Y&Y&Z&I&X&I&X&I&I&Y&Y&Y&I \\ \hline
\multirow{2}{*}{${\cal{S}}_I$}&X&X&Z&I&X&Z&Y&Y&Z&X&I&Z&X&X&I \\
&Z&Z&Y&I&Z&Y&X&X&Y&Z&I&Y&Z&Z&I\\
\hline\hline
\end{tabular}
\end{center}
\caption{Stabilizer generators of the [[15,9,3;$c=3$,$r=1$]] EAOQECC derived from the EAQECC given by Table~\ref{EAQECC2}.}
\label{EAOQECC2}
\end{table}

\begin{acknowledgments}
We wish to acknowledge enlightening discussions with David Poulin
and Graeme Smith.  TAB received financial support from
NSF Grant No.~CCF-0448658, and TAB and MHH both received support
from NSF Grant No.~ECS-0507270.  ID and MHH received financial
support from NSF Grant No.~CCF-0524811 and NSF Grant No.~CCF-0545845.
\end{acknowledgments}

\bibliography{ref4}

\end{document}